\documentclass[english,aps,prb,floatfix,twocolumn,showpacs,showkeys]{revtex4}
\usepackage[T1]{fontenc}
\usepackage[latin1]{inputenc}
\usepackage{amsmath}
\usepackage{babel}
\usepackage{graphicx}
\usepackage{amssymb}

\begin{document}
\title{Classical and {\it ab initio} preparation of reliable structures for
polymeric coordination compounds}

\author{Harald O. Jeschke}
\email{jeschke@itp.uni-frankfurt.de}
\affiliation{Institut für Theoretische Physik, Johann Wolfgang
 Goethe-Universität, Max-von-Laue-Str. 1, 60438 Frankfurt/Main, Germany}

\author{L. Andrea Salguero}
\affiliation{Institut für Theoretische Physik, Johann Wolfgang
 Goethe-Universität, Max-von-Laue-Str. 1, 60438 Frankfurt/Main, Germany}

\author{Roser Valentí}
\affiliation{Institut für Theoretische Physik, Johann Wolfgang
 Goethe-Universität, Max-von-Laue-Str. 1, 60438 Frankfurt/Main, Germany}

\author{Christian Buchsbaum}
\affiliation{Institut für Anorganische und Analytische Chemie, Johann Wolfgang
 Goethe-Universität, Max-von-Laue-Str. 7, 60438 Frankfurt/Main, Germany}

\author{Martin U. Schmidt}
\affiliation{Institut für Anorganische und Analytische Chemie, Johann Wolfgang
 Goethe-Universität, Max-von-Laue-Str. 7, 60438 Frankfurt/Main, Germany}

\author{Matthias Wagner}
\affiliation{Institut für Anorganische und Analytische Chemie, Johann Wolfgang
 Goethe-Universität, Max-von-Laue-Str. 7, 60438 Frankfurt/Main, Germany}

\date{\today}

\begin{abstract}
The detailed investigation of electronic and magnetic properties of
polymeric coordination materials with accurate {\it ab initio} quantum
mechanical methods is often computationally extremely demanding
because of the large number of atoms in the unit cell. Moreover,
usually the available structural data are insufficient or poorly
determined especially when the structure contains hydrogen atoms. In
order to be able to perform controlled {\it ab initio} calculations on
reliable structures, we use a two step approach to systematically
prepare model structures for polymeric coordination compound systems
and to relax them to their equilibrium configuration. First, a
structure is constructed on the basis of a crystallographic database
and optimized by force field methods; in the second step, the
structure is relaxed by {\it ab initio} quantum mechanical molecular
dynamics. With this structure we perform accurate electronic structure
calculations.  We will apply this procedure to a Fe(II) triazole
compound and to a coordination polymer of Cu(II) ions with
2,5-bis(pyrazol-1-yl)-1,4-dihydroxybenzene.
\end{abstract}




\maketitle

\section{Introduction}

Polymeric complexes form a class of materials that attract interest
from chemists and physicists. The possibility to apply organic
synthesis methods to produce families of compounds with similar
structures but subtly differing physical behavior offers an
opportunity to design materials with planned properties.

Materials showing temperature or pressure induced high-spin - low-spin
transitions are currently being discussed for applications in data
storage or spintronics. Such magnetic transitions can be optically
triggered, raising hopes for technological applications. The
cooperativity of such light-induced phase transitions also raises
questions concerning the underlying microscopic
processes~\cite{tayagaki:01,ogawa:00,koshihara:92,miyano:97}. Another
important class of polymeric coordination compounds are, in general,
low-dimensional magnetic materials where the coupling strengths
between the metal ions can be tuned by chemically modifying the
ligands and the transition metal ion coordination. Effects of, for
instance, dimensional crossover can be studied in this way.

Most of these compounds generally contain a large number of atoms in
the unit cell and a very low symmetry (in 80\% of the cases triclinic
or monoclinic) making the first principles study of the electronic and
magnetic properties of these systems under different conditions
(pressure, temperature, light irradiation) computationally very
demanding.  Moreover, since an accurate determination of hydrogen
positions with X-ray diffractometry is very difficult, usually the
available structure data are poorly determined implying unstable
quantum mechanical calculations. For some cases, the crystal
structures are not even known, due to absence of single crystals
suitable for an X-ray structure analysis. In order to overcome these
shortcomings, we use in this communication a two-step approach which
allows to perform accurate {\it ab initio} density functional theory
(DFT) calculations on reliable model structures for coordination
polymers. It should be noted, that the expression ``{\it ab initio}
prediction of crystal structures'' is also used in the sense of ``{\it
a priori} prediction'', i.e. predictions without reference to
diffraction data. Possible crystal structures of organic compounds
with 20--60 atoms can be predicted within some days or weeks using
force field methods~\cite{verwer:00}.  This works also for
organometallic compounds, e.g.  pentamethylferrocene
Fe(C$_5$H$_5$)[(C$_5$CH$_3$)$_5$]~\cite{schmidt:96}. Quantum
mechanical methods are much more time-consuming; hence their use is
limited to local optimizations starting from a given structural
model. Thus, quantum mechanical and force field methods can be
combined by making first a global optimization using force field
methods, followed by a local optimization with quantum mechanical
methods. This was done for many inorganic systems, e.g. NaCl or
MgF$_2$\cite{schoen:01}, and a few simple organic compounds,
e.g. glycole C$_2$H$_4$(OH)$_2$ and glycerol
C$_3$H$_5$(OH)$_3$~\cite{eijck:01}. In most other cases of crystal
structure predictions, quantum mechanical calculations were only used
for secondary calculations, e.g. to calculate molecular geometries and
electrostatic charges, or to derive intramolecular or intermolecular
potential curves for subsequent force field
optimizations~\cite{mooij:99}.

If experimental data are known, e.g. from crystal data of analogous
structures, this knowledge should of course be used as far as possible
in order to generate reliable model structures and to calculate solid
state properties. Here, we apply {\it ab initio} DFT methods to
understand the relationships between the structures and the electronic
and magnetic properties of both i) spin crossover materials and ii)
low-dimensional spin systems. We will consider two representatives
from this large class of compounds: i) a Fe(II)triazole
compound~\cite{kahn:98,garcia:02} which is thought to have a polymer
structure and ii) a coordination polymer of Cu(II) ions with
2,5-bis(pyrazol-1-yl)-1,4-dihydroxybenzene~\cite{dinnebier:02}, (we
shall denote it as [Cu(bpydhb)]$_n$, alternative names given in the
literature are CuCCP\cite{wolf:04}).

For the Fe(II)triazole system~\cite{kahn:98,garcia:02}, there is no
direct accurate determination of the crystal structure of the complex
as it cannot be crystallized sufficiently well. A chain-like structure
for such compounds of formula [Fe(Rtrz)$_3$]A$_2$.S where A denotes
the anion was predicted from EXAFS spectroscopy~\cite{Michalowicz:97}.
For analogous Cu(II) compounds this chain structure was confirmed by
single crystal X-ray analysis (see {\it e.g.} 
Refs.~\onlinecite{Drabent:01,garcia:97}). Polymeric iron triazole
compounds always show a poor crystallinity, and single crystals cannot
be grown. Hence the X-ray structure analyses are limited to molecular
compounds, e.g. trimers containing three Fe(II) ions connected by 6
triazole ligands and 6 terminal water
molecules~\cite{garcia:00}. Polymeric iron triazole complexes have
been studied by powder diffraction; but apart from a preliminary
indexing of [Fe(Htrz)$_3$](ClO$_4$)$_2\cdot1.85$ H$_2$O~\cite{smit:01}, no
structural data have been published so far. On the other hand, these
polymeric compounds show interesting high-spin--low-spin transitions
in the solid state (see {\it e.g.} Ref.~\onlinecite{guetlich:00}), and
quantum mechanical calculations could help to understand the mechanism
of this spin crossover.  Since all calculations of electronic and
magnetic properties require a reliable and reasonable model of the
crystal structure, we constructed the structures of a hypothetical
polymer crystal. All available information of molecular iron triazole
systems, and of analogous Cu(II) polymers was included. In order to
make the {\it ab initio} DFT calculations feasible we constructed at
first some models containing small triazoles (e.g. methyl-triazole)
and simple anions (e.g. F$^-$). Our rationale is that the model
compound still has the same coordination and immediate environment of
the Fe(II) ion as the real material which should thus allow us to
study the mechanisms of the high-spin - low-spin transition.
Structures with more complex anions will be considered later.

The second system that we study, the coordination polymer of Cu(II)
ions with 2,5-bis(pyrazol-1-yl)-1,4-dihydroxybenzene,
[Cu(bpydhb)]$_n$, has been recently synthesized by Wagner {\it et
al.}~\cite{dinnebier:02} The Cu$^{2+}$ ions in this compound form
spin-1/2 antiferromagnetic Heisenberg chains running along the
z-axis. We will analyze the influence on the electronic properties of
slight structural modifications, like substitutions of small organic
groups on the dihydroxybenzene or introduction of water ligands which
change the coordination of Cu from square to octahedral. The structure
of Cu(II) with 2,5-bis(pyrazol-1-yl)-1,4-dihydroxybenzene has been
obtained by X-ray powder diffraction~\cite{dinnebier:02} and we will
apply our two-step approach for the modified structures based on these
data.

\section{Methods}

We derive and investigate model polymeric coordination compound
structures in two steps. First, we create a rough structure on the
basis of crystallographic databases~\cite{database:05}. If no crystal
structures of similar compounds are known - like in the case of Fe
triazoles - we construct a hypothetical crystal structure having the
highest possible symmetry, a small number of atoms per unit cell and a
sensible arrangement of the individual fragments. The crystal
structure is optimized by force field methods including the
optimization of the unit cell parameters, but maintaining the
crystallographic symmetry. We used the Cerius$^2$ program
package~\cite{Cerius2} with a modified Dreiding~\cite{Mayo:90} force
field; atomic charges were calculated by the Gasteiger
method~\cite{Gasteiger:80}. Afterwards we proceed to improve the
structure by relaxing it again, but now with an {\it ab initio}
quantum mechanical molecular dynamics method~\cite{car:85}. Note that
we employ an all electron method already at this step in order to
obtain the best possible equilibrium structure even for demanding
transition metal complexes. Finally, we analyze the properties of the
relaxed model structure with an all electron {\it ab initio} DFT
method of the highest precision~\cite{wien}.

Our Car Parrinello (CP) {\it ab initio} molecular dynamics (AIMD)
calculations~\cite{car:85} are performed with a projector augmented
wave (PAW) basis set~\cite{bloechl:94}. This is an all electron
calculation based on the frozen core approximation. The
exchange-correlation potential is evaluated in the generalized
gradient approximation (GGA) as parametrized in
Ref.~\onlinecite{perdew:96}. After a first analysis of the electronic
properties for the classically prepared model in order to determine
the structural stability, we employ an AIMD with friction to refine
the structure. We impose constraints on the atomic coordinates in
order to preserve the space group symmetry of the compound.

For the final analysis step we employ the full potential linearized
augmented plane wave (LAPW, APW+lo) method as implemented in the
WIEN2k code~\cite{wien}.

\section{Results}

\subsection{Fe(II)-Triazole}

\begin{figure}
\includegraphics[angle=-90,width=0.45\textwidth]{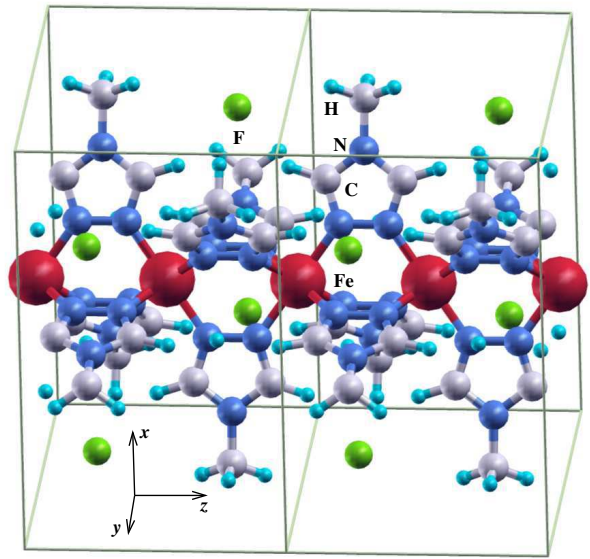}
\caption{Two unit cells of the triazole model
  polymer. The usually complicated anions have been replaced by
  single atom F$^-$ anions. }\label{fig:toymodel8}
\end{figure}

\begin{figure}

\includegraphics[angle=-90,width=0.45\textwidth]{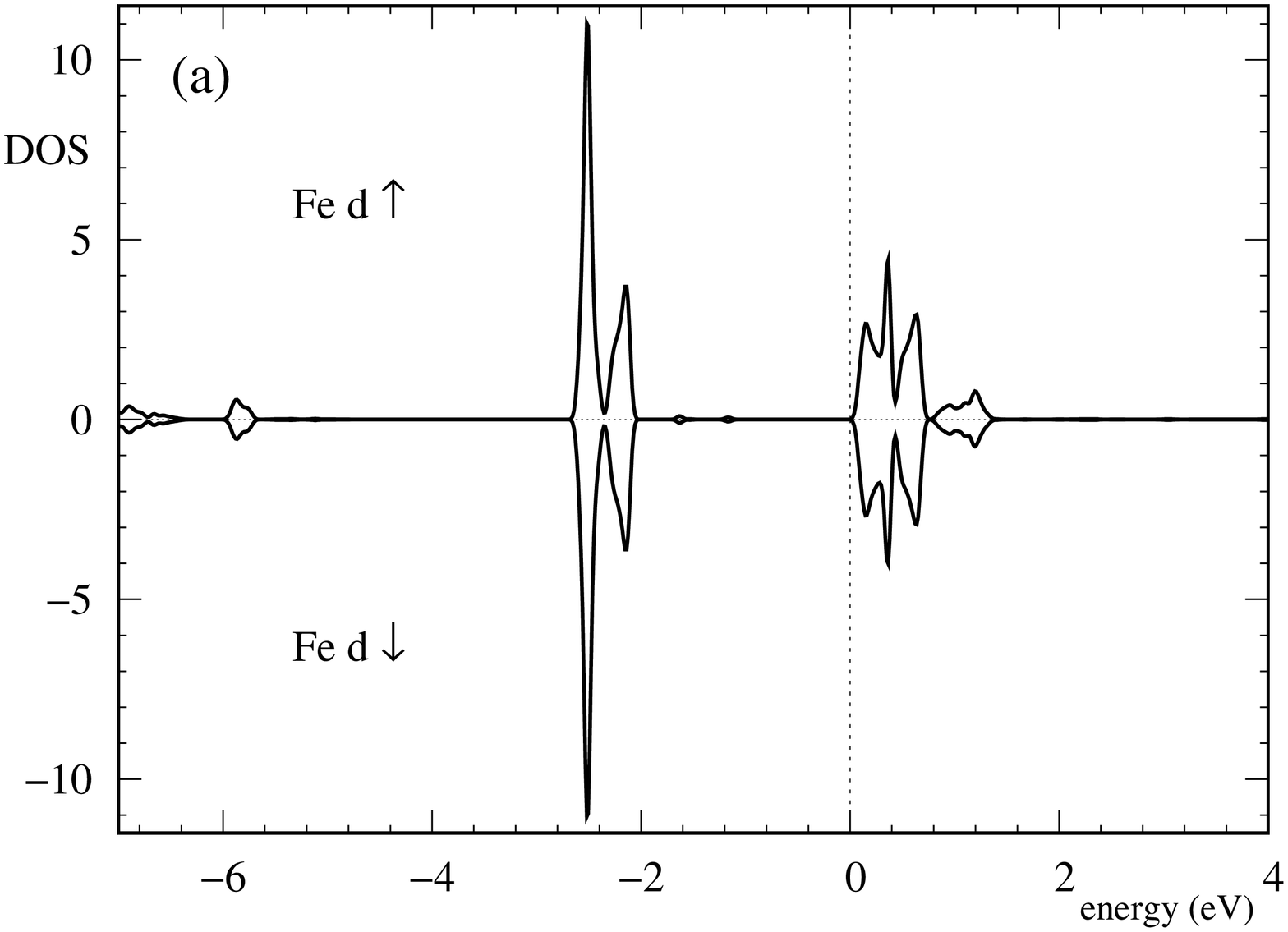}

\includegraphics[angle=-90,width=0.45\textwidth]{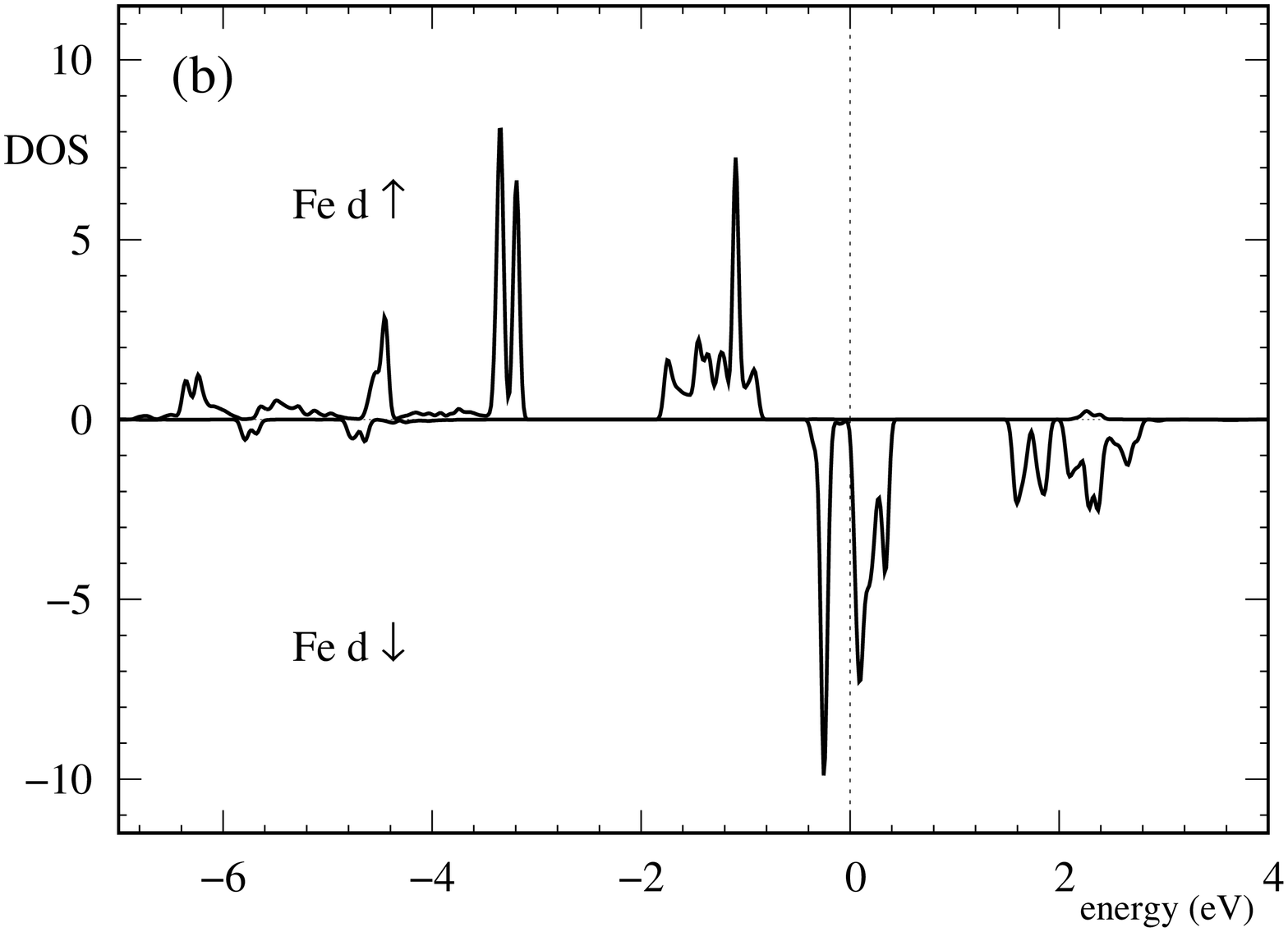}

\caption{Spin polarized Fe $d$ density of states of the Fe triazole
  structure Fig. \protect\ref{fig:toymodel8} in units of
  states/electronvolt/atom/spin. (a) for the S=0 state of Fe (b) for
  the S=2 state of Fe.  The upper and lower panels of the figure show
  respectively the density of states for the majority and minority
  spin states. The Fermi energy is indicated by a dotted line.
  }\label{fig:Fedos}
\end{figure}

The Fe(II) spin crossover molecular systems have been intensively
studied in the past~\cite{guetlich:00} and various theories have been
developed about the mechanism of the high-spin - low-spin
transition. A few years ago, Kahn {\it et al.}~\cite{kahn:98} proposed
that the cooperativity of this transition may be enhanced by replacing
molecular crystals by polymers, since in the latter the metallic sites
are bridged by chemical linkers through which one may get efficient
intersite coupling. A first principles DFT study of such systems would
bring some light into the properties of these
materials. Unfortunately, to date it has not been possible to get an
accurate X-ray structure determination of Fe(II) spin crossover
polymer systems, as discussed above, which is an indispensable input
for {\it ab initio} DFT calculations and therefore, to our knowledge,
no reliable {\it ab initio} studies were performed. With the two-step
method, we have obtained -- based on known data for molecular Fe(II)
spin crossover systems -- optimized model polymer structures which can
be used as input for {\it ab initio} DFT calculations. In
Fig.~\ref{fig:toymodel8} we present the structure of the polymer
Fe(II) compound with 4-methyl-1,2,4-triazole moieties as bridging
ligands between the Fe(II) cations. As anions we chose simple F$^-$
ions in order not to complicate the DFT calculations. The structure
has $P2_1/m$ symmetry; a higher symmetry was not possible with this
chemical composition. As an initial structure with low spin character,
the distances between the Fe(II) centers and the neighboring N atoms
of the triazole rings were chosen to be $d_\text{Fe-N}=1.97$~Å, close
to the Fe-N distance of $d_\text{Fe-N}=2.0$~Å that is known to be
typical for the low spin variants of Fe triazole
complexes\cite{Vos:83} while for the structure with high spin
character we chose $d_\text{Fe-N}=2.2$~Å . In Fig.~\ref{fig:Fedos} we
present the spin-polarized density of states (DOS)
obtained~\cite{wien} for this system after structural optimization for
the low spin (Fig.~\ref{fig:Fedos} (a)) and the high spin
(Fig.~\ref{fig:Fedos} (b)) states. Shown is the total Fe $d$ DOS for
majority (upper panel) and minority (lower panel) spin states. Since
Fe is in an octahedral environment of N, the 3 $d$ states split into
lower $t_{2g}$ $(d_{xy}, d_{xz}, d_{yz})$ and higher $e_g$
$(d_{x^2-y^2}, d_{z^2})$ states. In Fig.~\ref{fig:Fedos} (a) the 6
electrons of Fe(II) $3d^6$ occupy fully the $t_{2g}$ states (see DOS
manifold between -2 and -0.2 eV) for both spins thus corresponding to
a low spin S=0 state. The $e_g$ states are completely empty and lie
above the Fermi level.  In Fig.~\ref{fig:Fedos} (b), the majority
(spin up) $t_{2g}$ and $e_g$ states are occupied (upper panel DOS in
the range between -3.8eV and -0.6eV) while the minority (spin down,
lower panel) $t_{2g}$ states are about one-third occupied and
two-thirds empty. The minority $e_{g}$ states are completely empty
(states in the range of energies between 1eV and 3eV, lower panel).
These results indicate that the system is in a high spin S=2 state.  A
detailed investigation of the magnetic properties of this polymer
system for both spin states will be presented
elsewhere~\cite{paper_long}.

\subsection{Cu(II)-Polymer}

The Cu(II) coordination polymer [Cu(bpydhb)]$_n$  synthesized by Wagner {\it et
al.}~\cite{dinnebier:02} crystallizes in the triclinic space group $P
\bar{1}$ (No. 2) and consists of stacks of chains of deprotonated
2,5-bis(pyrazol-1-yl)-1,4-dihydroxybenzene molecules connected by
Cu$^{2+}$ ions. This system behaves as a homogeneous antiferromagnetic
spin-1/2 Heisenberg chain with an exchange coupling constant
$J=21.5$~K/k$_{\rm B}$ estimated from magnetic susceptibility
measurements~\cite{wolf:04}. The modular nature of this polymer makes
it an adequate candidate to study chemically feasible structural
modifications which may change the electronic and magnetic properties.
Of special interest is the effect of dimensionality, {\it i.e.} tuning
the crossover from one dimension to higher dimensions with the
inclusion of adequately modified ligands. Since in this communication
we want to stress the procedure we follow to design these structures
and how we calculate the electronic properties, we will concentrate
here on one substitution, namely two hydrogen atoms in the
hydroquinone ring by two amino groups and show the efficiency of the
two-step process. A second issue in this context that requires some
attention is the study of the effect of two additional ligands at the
Cu(II) ion, since such an octahedral coordination of copper is also
frequently observed. We will here present the relaxation process on
[Cu(bpydhb)]$_n$ with two additional water molecules in the
environment of copper.

\begin{figure}
\includegraphics[width=0.45\textwidth,angle=-0]{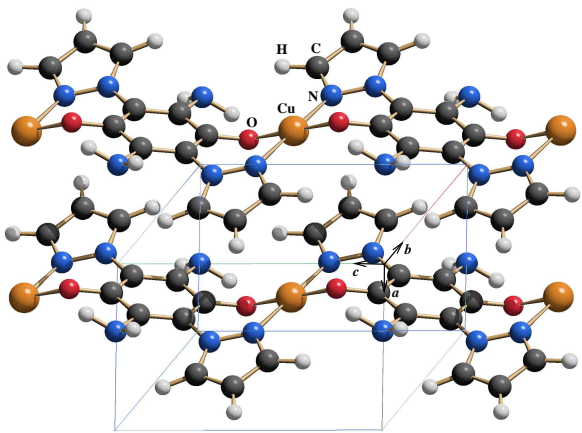}
\caption{Four unit cells of [Cu(bpydhb)]$_n$ (see text for the detailed
formula) where two hydrogen atoms in the hydroquinone rings have been
substituted by two amino groups. We find that the relaxed structure
shows some tilting of the amino groups out of the plane defined by the
benzene ring.}\label{fig:CuNH2}
\end{figure}

In Fig.~\ref{fig:CuNH2} we present the relaxed structure of
[Cu(bpydhb)]$_n$ where two hydrogen atoms in the hydroquinone rings
have been substituted by two amino groups. This structure was obtained
first by applying the force field method, which already has the
correct information of the chemical distances and angles from data
bases. A quantum mechanical evaluation of the forces on the atoms (see
Table ~\ref{tab:CuNH2force1}) for this structure, performed with the
full potential LAPW method~\cite{wien} (FPLAPW) still shows an
unstable structure. Our aim is to relax this structure showing forces
in the hundreds of mRyd/a$_{\rm B}$ with quantum mechanical molecular
dynamics calculations, AIMD, until we find forces smaller than
25~mRyd/a$_{\rm B}$ which we set as upper bound for the force
values. In order to get structures with forces smaller than
1~mRyd/a$_{\rm B}$ within the FPLAPW method, a fine tuning between the
various first principles DFT methods we employ has to be
considered. For our purposes here, a benchmark of 25~mRyd/a$_{\rm B}$
is reasonable.

In Table~\ref{tab:CuNH2force2} the forces after the AIMD relaxation
are shown and we observe that the structure is relaxed to our required
precision. The fractional coordinates of this structure are given in
Table~\ref{tab:CuNH2pos} where the symmetry and cell parameters were
kept unchanged with respect to the unsubstituted Cu polymer system
[Cu(bpydhb)]$_n$~\cite{dinnebier:02} $a=5.1723$~Å, $b=7.9587$~Å,
$c=8.2298$~Å, and angles $\alpha=118.221$°, $\beta=91.520$°, $\gamma=100.148$°.

\begin{table}
  \caption{ Forces~\protect\cite{wien} between the atoms for the
  Cu(II)-NH$_{2}$ polymer structure designed with the force field
  method. Note that the forces were determined under the constraint of
  inversion symmetry; thus there are no forces acting on the Cu(II)
  ion center. Units are in mRyd/a$_{\rm B}$. See
  Table~\protect\ref{tab:CuNH2pos} for identification of the atom
  location. }
\begin{tabular}{crrr} \hline
Atom & F$_{x}$ & F$_{y}$ & F$_{z}$ \\ \hline \hline
Cu & $0$ & $0$ & $0$ \\
O & $-47.4$ & $-62.7$ & $51.3$ \\
N1 & $89.1$ & $32.8$ & $-42.6$ \\
N2 & $22.4$ & $27.2$ & $-95.8$ \\
N3 & $-94.2$ & $-66.8$ & $33.5$ \\
C1 & $35.0$ & $22.9$ & $1.8$ \\
C2 & $-92.7$ & $-29.2$ & $111.1$ \\
C3 & $-39.2$ & $-41.7$ & $294.8$ \\
C4 & $-239.7$ & $-104.1$ & $16.5$ \\
C5 & $-26.6$ & $-12.8$ & $-304.2$ \\
C6 & $106.7$ & $88.9$ & $46.0$ \\
H1 & $128.6$ & $-28.7$ & $215.0$ \\
H2 & $255.3$ & $106.1$ & $-34.2$ \\
H3 & $23.8$ & $25.6$ & $-279.0$ \\
H4 & $4.8$ & $10.8$ & $-18.3$ \\
H5 & $-16.2$ & $64.5$ & $56.3$ \\ \hline

\end{tabular}
\label{tab:CuNH2force1}
\end{table}

\begin{table}
  \caption{Forces\protect\cite{wien} between the atoms for the {\it
  relaxed} Cu(II)-NH$_{2}$ polymer. Units are mRyd/a$_{\rm B}$.}
\begin{tabular}{crrr} \hline
Atom & F$_{x}$ & F$_{y}$ & F$_{z}$ \\ \hline \hline
Cu & $0$ & $0$ & $0$ \\
O & $ 12.7$ & $16.9$ & $21.7$ \\
N1 & $ 3.9$ & $0.7$ & $-17.2$ \\
N2 & $ 12.5$ & $-2.3$ & $8.4$ \\
N3 & $ -21.0$ & $-9.4$ & $-5.5$ \\
C1 & $ -1.0$ & $-9.3$ & $-19.4$ \\
C2 & $10.3$ & $3.6$ & $-2.1  $ \\ 
C3 & $-7.1$ & $-7.0$ & $14.6 $ \\ 
C4 & $ -10.6$ & $-12.7$ & $-3.1$ \\
C5 & $ -12.0$ & $-6.0$ & $-19.5$ \\ 
C6 & $ 12.0$ & $3.4$ & $-0.7  $ \\ 
H1 & $ 1.4$ & $-0.7$ & $3.5$ \\ 
H2 & $ 3.1$ & $-1.7$ & $-1.4$ \\ 
H3 & $ -0.4$ & $-0.3$ & $-4.1$ \\ 
H4 & $0.08$ & $-5.5$ & $ -0.4$ \\ 
H5 & $1.1$ & $-0.9$ & $0.5$ \\ \hline 
\end{tabular}
\label{tab:CuNH2force2}
\end{table}

\begin{table}
\caption{
Fractional atomic positions of nonequivalent atoms in Cu(II)-NH$_{2}$
obtained after relaxation. The corresponding unit cell parameters are
given in the text.}
\begin{tabular}{cccc} \hline
Atom & $x$ & $y$ & $z$ \\ \hline \hline
Cu & $0.5$ & $0.5$ & $0.5 $ \\
O & $0.45952338$ & $0.34384463$ & $0.62047714$ \\
N1 & $0.87129861$ & $0.67649196$ & $0.85703667$ \\
N2 & $0.81592808$ & $0.67403219$ & $0.69273582$ \\
N3 & $0.89342131$ & $0.81953377$ & $0.25313235$ \\
C1 & $ 0.48379900$ & $0.42607948$ & $0.80332348$ \\
C2 & $ 0.68236558$ & $0.58873536$ & $0.92870320$ \\
C3 & $ 0.03998649$ & $0.76736537$ & $0.66677421$ \\
C4 & $ 0.23751744$ & $0.83563448$ & $0.81694695$ \\
C5 & $ 0.12514376$ & $0.77281287$ & $0.93447831$ \\
C6 & $ 0.70113966$ & $0.66249861$ & $0.12137891$ \\
H1 & $ 0.20669437$ & $0.78002257$ & $0.06146726$ \\
H2 & $ 0.43843580$ & $0.91660682$ & $0.83522646$ \\
H3 & $ 0.04936408$ & $0.77481595$ & $0.53807198$ \\
H4 & $ 0.82019516$ & $0.86915420$ & $0.37967013$ \\
H5 & $ 0.94873662$ & $0.93065156$ & $0.22239484$ \\ \hline

\end{tabular}
\label{tab:CuNH2pos}
\end{table}

\begin{figure}
\includegraphics[width=0.45\textwidth,angle=-0]{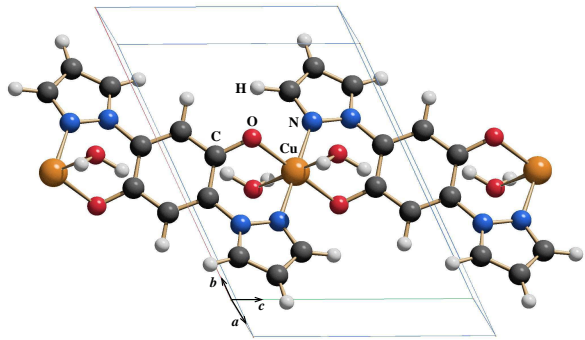}
\caption{One unit cell of the  [Cu(bpydhb)]$_n$  polymer with water molecules
  bonding to the Cu(II) ion. The unit cell has been relaxed by the
  force field method but kept fixed in the AIMD relaxation. It differs
  considerably from the original [Cu(bpydhb)]$_n$ polymer unit
  cell. The equilibrium positions of the water molecule lead to a
  nearly octahedrally coordinated Cu(II) ion.}\label{fig:CuH2O}
\end{figure}

In a different modification of the Cu(II) polymer structure, we want
to investigate the influence of a change of the Cu coordination from
four to six on the electronic properties of the Cu(II) polymer. We
added two water molecules bound to the Cu center and proceeded as
outlined for the case of the NH$_2$ substitution: A first structure
was obtained from the force field method, yielding high forces when we
controlled it with an FPLAPW calculation. Further relaxation with AIMD
yields forces around 20~mRyd/a$_{\rm B}$. The corresponding structure
is shown in Fig.~\ref{fig:CuH2O}, and the fractional coordinates are
given in Table~\ref{tab:CuH2Opos}. The new unit cell is triclinic with
space group $P\bar{1}$ (No. 2) as in the original Cu(II) polymer. The
cell parameters are $a=5.2341$~Å, $b=11.2493$~Å, $c=8.0721$~Å and the
angles are $\alpha=117.6108$°, $\beta=68.8218$°, $\gamma=127.1551$°. Note the
variation of these parameters with respect to those of
[Cu(bpydhb)]$_n$~\cite{dinnebier:02}, since in the relaxation process
of the new structure the unit cell was allowed to relax properly in
order to accommodate the water molecules.

\begin{table}
  \caption{ Fractional atomic positions of nonequivalent atoms in
  Cu(II)-H$_{2}$O obtained after relaxation. The corresponding unit
  cell parameters are given in the text.}
\begin{tabular}{cccc} \hline
Atom & $x$ & $y$ & $z$ \\ \hline \hline
Cu & $0.5$ & $0.5$ & $0.5 $ \\
O1 & $0.54548630$ & $0.39146013$ & $0.61449576$ \\ 
O2 & $0.11104809$ & $0.50997119$ & $0.71612732$ \\ 
C1 & $0.50846703$ & $0.44754970$ & $0.80067207$ \\ 
C2 & $0.68032852$ & $0.61183880$ & $0.90724635$ \\ 
C3 & $0.05894527$ & $0.84496851$ & $0.61704992$ \\ 
C4 & $0.26950758$ & $0.95163087$ & $0.75697623$ \\ 
C5 & $0.15279255$ & $0.87395244$ & $0.88376263$ \\ 
C6 & $ 0.32801067$ & $0.33993413$ & $0.89980571$ \\  
N1 & $0.88662263$ & $0.72926141$ & $0.81944391$ \\ 
N2 & $0.82847386$ & $0.71119262 $ & $0.65595860$ \\ 
H1 & $0.23644016$ & $0.90962503$ & $0.01364916$ \\ 
H2 & $0.47761970$ & $0.07047826$ & $0.76570604$ \\ 
H3 & $ 0.05611367$ & $0.85699006$ & $0.49132676$ \\ 
H4 & $0.20991897$ & $0.21421176$ & $0.81558567$ \\
H5 & $0.02924495$ & $0.44665601$ & $0.79637256$ \\
H6 & $0.92725690$ & $0.45505960$ & $0.64249638$ \\ \hline

\end{tabular}
\label{tab:CuH2Opos}
\end{table}

\begin{figure}
\includegraphics[width=0.45\textwidth,angle=-90]{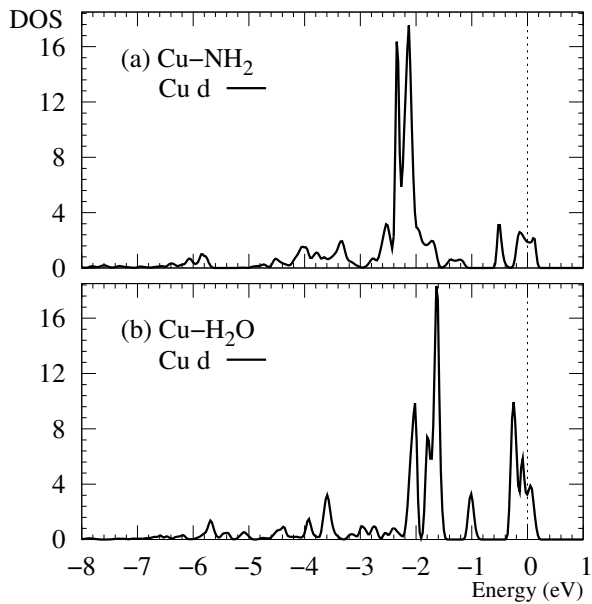}
\caption{(a) Cu $d$ density of states of the Cu polymer [Cu(bpydhb)]$_n$ with
  two NH$_2$ groups, and (b) of the Cu polymer [Cu(bpydhb)]$_n$ with
  two water ligands.}\label{fig:Cudos}
\end{figure}

We investigate now the electronic properties of the two model Cu(II)
polymers with a precise FPLAPW calculation. A detailed account of the
electronic and magnetic properties of both systems is presented
elsewhere~\cite{paper_long2}. Here we would like to stress the effect
of a planar Cu environment (for the Cu-NH$_2$ case) in comparison of
an octahedral environment (for the Cu-H$_2$O case) on the Cu $d$
density of states. In Figs.~\ref{fig:Cudos}~(a) and (b) the Cu $d$
density of states for both modifications is presented. While for both
cases Cu is in an oxidation $2+$ with the $d_{x^2-y^2}$ orbital half
filled, the effect of the octahedral crystal field shows a splitting
into $t_{2g}$ ($d_{xy}, d_{xz}, d_{yz}$) for energies below -1.5~eV
and $e_g$ $(d_{z^2}, d_{x^2-y^2})$ states above -1.5~eV (see
Fig.~\ref{fig:Cudos}~(b)) while in the planar environment all $d$
states are split and the states contributing to the bonding at the
Fermi surface
are mainly of $d_{x^2-y^2}$ character in the local coordinate frame of
Cu which is defined with the local $z$ direction pointing from the Cu
to the O of the H$_2$O molecule and the $y$ direction pointing from
the Cu to the in-plane O atom.

\section{Conclusions}

We have investigated the structural properties and its effects on the
electronic properties of the quasi-one-dimensional polymer
[Cu(bpydhb)]$_n$ with small structural modifications and a model
Fe(II) triazole polymer using classical and {\it ab initio} DFT
methods. We demonstrate that the use of a stepwise approach to the
study of complicated coordination polymer materials is an effective
procedure to obtain reliable structures for accurate quantum
mechanical analysis. We note that an exclusive use of {\it ab initio}
molecular dynamics is very time consuming while the combination of the
classical with the quantum approaches speeds up the relaxation process
considerably. We first employ a classical force field to create and
relax plausible initial structures. In a second step, we proceed with
{\it ab initio} molecular dynamics to relax these structures to a
stable configuration. Finally we extract the electronic properties of
the transition metal compounds with precise FPLAPW calculations.

\begin{acknowledgments}
H.O.J. gratefully acknowledges support from the DFG through the Emmy
Noether Program.  This work was financially supported by the Deutsche
Forschungsgemeinschaft under the auspices of the Forschergruppe 412 on
{\it ``spin and charge correlations in low-dimensional metalorganic
solids''}. We thank A. Kokalj for providing the visualization code
XcrySDen~\cite{kokalj:03}. We gratefully acknowledge support by the
Frankfurt Center for Scientific Computing.
\end{acknowledgments}

\end{document}